\RequirePackage{lineno}
\documentclass[aps,prd,twocolumn,superscriptaddress,showpacs]{revtex4}
\usepackage{epsfig}
\newcommand{\syspscale}{17}
\newcommand{\syslepresmte}{9}
\newcommand{\syslepresmtm}{3}

\newcommand{\sysqedmte}{11}
\newcommand{\sysqedmtm}{12}

\newcommand{\sysrecrespmt}{9}

\newcommand{\sysrecresolmt}{7}

\newcommand{\sysgtwomt}{3}

\newcommand{\sysescale}{30}
\newcommand{\sysmbkgmt}{9}

\def\met{\mbox{${\hbox{$E$\kern-0.6em\lower-.1ex\hbox{/}}}_T$}} %missing ET
\def\mex{\mbox{${\hbox{$E$\kern-0.6em\lower-.1ex\hbox{/}}}_x$}} %missing Ex
\def\mey{\mbox{${\hbox{$E$\kern-0.6em\lower-.1ex\hbox{/}}}_y$}} %missing Ey
\def\mexy{\mbox{${\hbox{$E$\kern-0.6em\lower-.1ex\hbox{/}}}_{x,y}$}} %missing Exy
 
\def\mevcc{MeV/$c^2$}                   %MeV/c^2
\def\gevcc{GeV/$c^2~$}                   %GeV/c^2
\def\gevc{GeV/$c~$}                       %GeV/c
\def\D0{D\O}                            %D0
\begin{document}
\lefthyphenmin=2
\righthyphenmin=3

%new 060600
%\setlength{\baselineskip}{24pt}
%\preprint{FERMILAB-Pub-97/109-E}
%
% ======> Title of the paper goes here <====================
%

\title{First Measurement of the $W$ Boson Mass \\ 
 in Run II of the Tevatron }% Force line breaks with \\

\affiliation{Institute of Physics, Academia Sinica, Taipei, Taiwan 11529, Republic of China} 
\affiliation{Argonne National Laboratory, Argonne, Illinois 60439} 
\affiliation{Institut de Fisica d'Altes Energies, Universitat Autonoma de Barcelona, E-08193, Bellaterra (Barcelona), Spain} 
\affiliation{Baylor University, Waco, Texas  76798} 
\affiliation{Istituto Nazionale di Fisica Nucleare, University of Bologna, I-40127 Bologna, Italy} 
\affiliation{Brandeis University, Waltham, Massachusetts 02254} 
\affiliation{University of California, Davis, Davis, California  95616} 
\affiliation{University of California, Los Angeles, Los Angeles, California  90024} 
\affiliation{University of California, San Diego, La Jolla, California  92093} 
\affiliation{University of California, Santa Barbara, Santa Barbara, California 93106} 
\affiliation{Instituto de Fisica de Cantabria, CSIC-University of Cantabria, 39005 Santander, Spain} 
\affiliation{Carnegie Mellon University, Pittsburgh, PA  15213} 
\affiliation{Enrico Fermi Institute, University of Chicago, Chicago, Illinois 60637} 
\affiliation{Comenius University, 842 48 Bratislava, Slovakia; Institute of Experimental Physics, 040 01 Kosice, Slovakia} 
\affiliation{Joint Institute for Nuclear Research, RU-141980 Dubna, Russia} 
\affiliation{Duke University, Durham, North Carolina  27708} 
\affiliation{Fermi National Accelerator Laboratory, Batavia, Illinois 60510} 
\affiliation{University of Florida, Gainesville, Florida  32611} 
\affiliation{Laboratori Nazionali di Frascati, Istituto Nazionale di Fisica Nucleare, I-00044 Frascati, Italy} 
\affiliation{University of Geneva, CH-1211 Geneva 4, Switzerland} 
\affiliation{Glasgow University, Glasgow G12 8QQ, United Kingdom} 
\affiliation{Harvard University, Cambridge, Massachusetts 02138} 
\affiliation{Division of High Energy Physics, Department of Physics, University of Helsinki and Helsinki Institute of Physics, FIN-00014, Helsinki, Finland} 
\affiliation{University of Illinois, Urbana, Illinois 61801} 
\affiliation{The Johns Hopkins University, Baltimore, Maryland 21218} 
\affiliation{Institut f\"{u}r Experimentelle Kernphysik, Universit\"{a}t Karlsruhe, 76128 Karlsruhe, Germany} 
\affiliation{High Energy Accelerator Research Organization (KEK), Tsukuba, Ibaraki 305, Japan} 
\affiliation{Center for High Energy Physics: Kyungpook National University, Taegu 702-701, Korea; Seoul National University, Seoul 151-742, Korea; SungKyunKwan University, Suwon 440-746, Korea} 
\affiliation{Ernest Orlando Lawrence Berkeley National Laboratory, Berkeley, California 94720} 
\affiliation{University of Liverpool, Liverpool L69 7ZE, United Kingdom} 
\affiliation{University College London, London WC1E 6BT, United Kingdom} 
\affiliation{Centro de Investigaciones Energeticas Medioambientales y Tecnologicas, E-28040 Madrid, Spain} 
\affiliation{Massachusetts Institute of Technology, Cambridge, Massachusetts  02139} 
\affiliation{Institute of Particle Physics: McGill University, Montr\'{e}al, Canada H3A~2T8; and University of Toronto, Toronto, Canada M5S~1A7} 
\affiliation{University of Michigan, Ann Arbor, Michigan 48109} 
\affiliation{Michigan State University, East Lansing, Michigan  48824} 
\affiliation{University of New Mexico, Albuquerque, New Mexico 87131} 
\affiliation{Northwestern University, Evanston, Illinois  60208} 
\affiliation{The Ohio State University, Columbus, Ohio  43210} 
\affiliation{Okayama University, Okayama 700-8530, Japan} 
\affiliation{Osaka City University, Osaka 588, Japan} 
\affiliation{University of Oxford, Oxford OX1 3RH, United Kingdom} 
\affiliation{University of Padova, Istituto Nazionale di Fisica Nucleare, Sezione di Padova-Trento, I-35131 Padova, Italy} 
\affiliation{LPNHE, Universite Pierre et Marie Curie/IN2P3-CNRS, UMR7585, Paris, F-75252 France} 
\affiliation{University of Pennsylvania, Philadelphia, Pennsylvania 19104} 
\affiliation{Istituto Nazionale di Fisica Nucleare Pisa, Universities of Pisa, Siena and Scuola Normale Superiore, I-56127 Pisa, Italy} 
\affiliation{University of Pittsburgh, Pittsburgh, Pennsylvania 15260} 
\affiliation{Purdue University, West Lafayette, Indiana 47907} 
\affiliation{University of Rochester, Rochester, New York 14627} 
\affiliation{The Rockefeller University, New York, New York 10021} 
\affiliation{Istituto Nazionale di Fisica Nucleare, Sezione di Roma 1, University of Rome ``La Sapienza," I-00185 Roma, Italy} 
\affiliation{Rutgers University, Piscataway, New Jersey 08855} 
\affiliation{Texas A\&M University, College Station, Texas 77843} 
\affiliation{Istituto Nazionale di Fisica Nucleare, University of Trieste/\ Udine, Italy} 
\affiliation{University of Tsukuba, Tsukuba, Ibaraki 305, Japan} 
\affiliation{Tufts University, Medford, Massachusetts 02155} 
\affiliation{Waseda University, Tokyo 169, Japan} 
\affiliation{Wayne State University, Detroit, Michigan  48201} 
\affiliation{University of Wisconsin, Madison, Wisconsin 53706} 
\affiliation{Yale University, New Haven, Connecticut 06520} 
\author{T.~Aaltonen}
\affiliation{Division of High Energy Physics, Department of Physics, University of Helsinki and Helsinki Institute of Physics, FIN-00014, Helsinki, Finland}
\author{A.~Abulencia}
\affiliation{University of Illinois, Urbana, Illinois 61801}
\author{J.~Adelman}
\affiliation{Enrico Fermi Institute, University of Chicago, Chicago, Illinois 60637}
\author{T.~Affolder}
\affiliation{University of California, Santa Barbara, Santa Barbara, California 93106}
\author{T.~Akimoto}
\affiliation{University of Tsukuba, Tsukuba, Ibaraki 305, Japan}
\author{M.G.~Albrow}
\affiliation{Fermi National Accelerator Laboratory, Batavia, Illinois 60510}
\author{S.~Amerio}
\affiliation{University of Padova, Istituto Nazionale di Fisica Nucleare, Sezione di Padova-Trento, I-35131 Padova, Italy}
\author{D.~Amidei}
\affiliation{University of Michigan, Ann Arbor, Michigan 48109}
\author{A.~Anastassov}
\affiliation{Rutgers University, Piscataway, New Jersey 08855}
\author{K.~Anikeev}
\affiliation{Fermi National Accelerator Laboratory, Batavia, Illinois 60510}
% ...
\collaboration{CDF Collaboration\footnote{
With visitors from $^a$University of Athens, 15784 Athens, Greece, 
$^b$University of Bristol, Bristol BS8 1TL, United Kingdom, 
$^c$University Libre de Bruxelles, B-1050 Brussels, Belgium, 
$^d$Cornell University, Ithaca, NY  14853, 
$^e$University of Cyprus, Nicosia CY-1678, Cyprus, 
$^f$University College Dublin, Dublin 4, Ireland, 
$^g$University of Edinburgh, Edinburgh EH9 3JZ, United Kingdom, 
$^h$University of Heidelberg, D-69120 Heidelberg, Germany, 
$^i$Universidad Iberoamericana, Mexico D.F., Mexico, 
$^j$University of Manchester, Manchester M13 9PL, England, 
$^k$Nagasaki Institute of Applied Science, Nagasaki, Japan, 
$^l$University de Oviedo, E-33007 Oviedo, Spain, 
$^m$University of London, Queen Mary College, London, E1 4NS, England, 
$^n$University of California Santa Cruz, Santa Cruz, CA  95064, 
$^o$Texas Tech University, Lubbock, TX  79409, 
$^p$University of California, Irvine, Irvine, CA  92697, 
$^q$IFIC(CSIC-Universitat de Valencia), 46071 Valencia, Spain.}
}
\noaffiliation

%
% ==============> Text of the abstract goes here <=====================
% 
\begin{abstract}
 We present a measurement of the $W$ boson mass using 200  pb$^{-1}$  of data collected in 
 {\mbox{$p\bar p$}}\ collisions at {\mbox{$\sqrt{s}$ =\ 1.96\ TeV}} by
 the CDF II detector at Run II of the Fermilab Tevatron. With a  sample of 63964 $W \rightarrow 
 e \nu$ candidates and 51128 $W \rightarrow \mu \nu$ candidates, we measure $M_W = ( 80413 \pm
 34_{\rm stat} \pm 34_{\rm syst}  =   80413 \pm 48 ) $ \mevcc. This is
 the most precise single measurement of the $W$ boson mass to date.  
\end{abstract}


\maketitle

%\newpage
\vfill\eject
%\twocolumn

The standard model (SM) invokes the Higgs mechanism of spontaneous symmetry breaking
 to generate mass for the $W$ and $Z$ bosons, which mediate the weak force.
 The $SU(2) \times U(1)$ symmetry of the electroweak interaction predicts
 the relation
 between the $W$ and $Z$ boson masses and the electromagnetic and weak gauge couplings.
 The prediction for the $W$ boson mass $M_W$ in terms of the precisely measured
 $Z$ boson mass $M_Z$, the Fermi decay constant $G_F$ extracted from the muon
 lifetime measurement, and the electromagnetic coupling $\alpha$ at the 
 scale $M_Z$, is given in the 
 ``on-shell'' scheme by~\cite{langacker}
$$ 
M_W^2 = \frac{\hbar^3}{c}\frac { \pi \alpha} {\sqrt{2} G_F } \frac{1}{(1-c^2_W) (1 - \Delta r)}  \quad,
$$
 where $c_W$ = $M_W / M_Z$ and $\Delta r$ is the quantum-loop correction. 
 A precise measurement of $M_W$ provides a measurement of $\Delta r$.
 In the SM the contributions to $\Delta r$
 are dominated by the top quark and the Higgs boson loops, such that  $M_W$ in conjunction with
 the top quark mass constrains the mass $m_H$ of the undiscovered Higgs boson. 
 An $m_H$ constraint inconsistent with 
 direct searches can indicate the presence of new physics, such as
 contributions to $\Delta r$ from supersymmetric particles~\cite{heinemeyer}. 

The $W$ boson mass~\cite{langacker} has been measured most precisely  
 by the  LEP~\cite{LEPexp,lepwmass} and 
 Tevatron~\cite{cdfd0} experiments, with the world-average
 $M_W = ( 80392 \pm 29 ) $~\mevcc~\cite{lepwmass}.  At the Tevatron, 
 $W$ bosons are mainly produced in quark ($q^\prime$) anti-quark ($\bar{q}$) annihilation 
 $q^\prime \bar{q} \rightarrow W + X $. Here $X$ includes the 
 QCD radiation that forms the ``hadronic recoil'' balancing
 the boson's transverse momentum $p_T$~\cite{coords}. The 
 $W \rightarrow \ell \nu$ decays,
 characterized by a high-$p_T$ charged lepton ($\ell = e \; \mathrm{or} \; \mu$) and neutrino, 
 can be selected with high purity and 
 provide precise mass information. 

 This analysis~\cite{prd,oliver}  uses 200~pb$^{-1}$  collected by the CDF II detector~\cite{prd}
  in $p \bar{p} $ collisions at $\sqrt{s} = 1.96$~TeV at the 
 Tevatron. CDF II is a magnetic spectrometer surrounded by calorimeters
 and muon detectors. We use the 
  central drift chamber
  (COT)~\cite{cot},  the central 
 calorimeter~\cite{run1detector} with embedded wire 
  chambers~\cite{ces} at the electromagnetic (EM) shower maximum, and the muon 
 detectors~\cite{muondetectors}  
 for identification of muons and electrons with 
 $| \eta | < 1$~\cite{coords} and measurement 
of their four-momenta. 
 The muon (electron) trigger requires a COT track with 
  $p_T > 18 (9) $~\gevc\ \cite{coords}, and 
 matching muon chamber hits  (EM calorimeter cluster with  $E_T > 18 $~GeV). 

 In the  analysis, we select muons
 with  a COT track matched to  muon chamber hits and passing quality requirements, track $p_T > 30$~\gevc, and 
 a minimum-ionization signal in the calorimeter. Cosmic rays are rejected
  using  COT
  hit timing~\cite{cotcosmic}. We select electrons with track $p_T > 18$~\gevc,
 EM cluster $E_T > 30 $~GeV~\cite{prd,coords}, and passing quality requirements on the COT track and the track-cluster
 matching. Additional 
 requirements are based on the ratio of the calorimeter energy $E$ to track momentum $p$ ($E / pc < 2$), 
 the ratio of energies  detected in the hadronic and EM calorimeters $E_{\mathit{Had}}/E_{\mathit{EM}} < 0.1$, 
 and the transverse shower profile~\cite{prd}. A veto 
 on the presence of a second lepton suppresses $Z$ boson background,
 with negligible loss of $W$ boson events. 
 Control samples of $Z$ boson events require 
 two oppositely charged leptons with the above criteria. 

 The  $\vec{p}_T$ of the hadronic recoil  
 ($\vec{u}$) equals the vector sum
 $\vec{u} = \Sigma_i  E_i \sin (\theta_i)  \hat{n}_i / c$
  over calorimeter towers~\cite{run1detector}, 
 with energy $E_i$, polar angle $\theta_i$, and 
 transverse directions specified by unit vectors $\hat{n}_i$. Energy associated with the charged lepton(s) is not included.
 We impose $\vec{p}_T$ balance
 to infer the neutrino's transverse momentum
  $p_T^\nu \equiv | - \vec{p}_T^{~\ell} - \vec{u} |$~\cite{coords} and
 the $W$ transverse mass $m_T = \sqrt {
 2 ~ (~p_T^\ell ~ p_T^\nu  - \vec{p}_T^{~\ell} \cdot \vec{p}_T^{~\nu}) }/c$. 
  We require $p_T^\nu > 30$~\gevc\ and $|\vec{u}| < 15$~\gevc\ to obtain a $W$ candidate 
 sample of high purity, whose $m_T$ and lepton 
 $p_T$  distributions are strongly correlated with $M_W$. 
 The sample consists of
  63964 $W \rightarrow
 e \nu$  and 51128 $W \rightarrow \mu \nu$ candidates. 

 The $W$ boson mass is extracted by performing binned maximum likelihood fits to the distributions
 of $m_T$, $p_T^\ell$  and $p_T^\nu$. We generate 800 
 templates  as functions of $M_W$ between 80 \gevcc and 81 \gevcc 
 using a custom Monte Carlo (MC)
 simulation~\cite{prd}  
 of boson production and decay, and of the detector response to the 
 lepton(s) and hadronic recoil. The custom MC optimizes computing speed and control
 of systematic uncertainties. 
 The  kinematics of $W$ and $Z$ boson decays are obtained
 from the {\sc resbos}~\cite{resbos} program. 
 We tune the non-perturbative form factor in {\sc resbos}, which describes the boson $p_T$ spectrum at
 low $p_T$,  on the dilepton $p_T$ distributions in the $Z$ boson data.
  Single photons (FSR) radiated from the final-state leptons 
 are generated according to the {\sc wgrad}
  program~\cite{wgrad}.   The FSR photon energies are increased by 10\% (with an absolute 
 uncertainty of 5\%) to account for
  additional energy loss due to 
 two-photon radiation~\cite{carloni}.  {\sc wgrad} is also used to estimate the 
   initial-state QED radiation.
 We use the CTEQ6M~\cite{cteq6}
 set of parton distribution functions and  their uncertainties.

 The custom MC performs a hit-level simulation
 of the lepton track. A 
 fine-grained model of passive material properties 
 is used   
  to calculate ionization and radiative energy loss
 and multiple Coulomb scattering. Bremsstrahlung photons
 and conversion electrons  are
 generated and propagated to the calorimeter. COT
 hits are generated  according to the  
 resolution ($\approx 150 ~ \mu$m) and efficiencies measured from
 muon tracks in $\Upsilon, W$, and $Z$ boson decays. A helix 
 fit (with optional beam constraint) is performed to simulate the reconstructed track. 

  The  alignment of the COT  is performed using a high-purity sample of high-$p_T$
 cosmic ray muons. Each muon's complete trajectory  is
 fit to a single helix~\cite{cotcosmic}. The fits determine  the  relative locations of the sense wires, including
 gravitational and electrostatic displacements, with a precision of a few microns. We constrain remaining misalignments,
 which cause a bias in the track curvature, by comparing  $\left\langle E/pc \right\rangle$ for electrons
 and positrons. 

\begin{figure}[htb]
\begin{center}
\epsfysize = 3.9cm
\epsffile {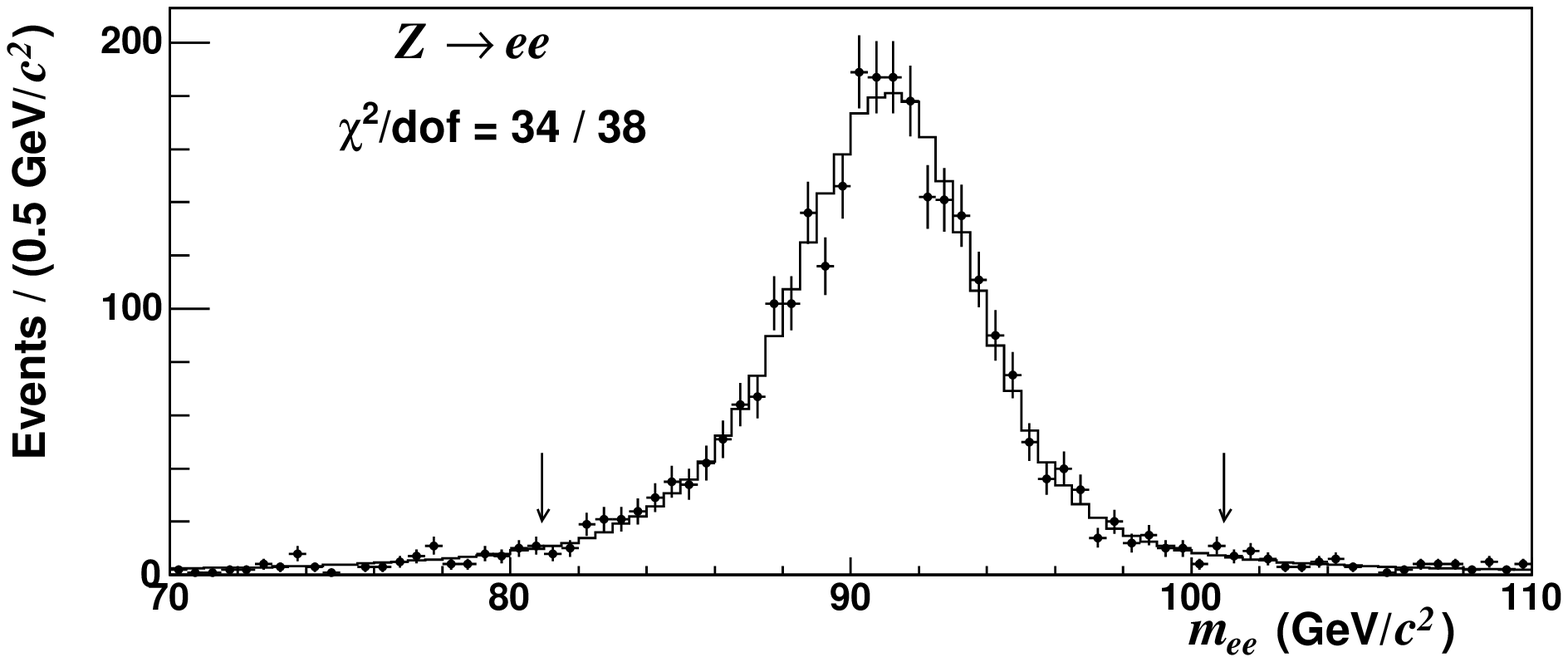}
\epsfysize = 3.9cm
\vspace*{-4mm}
\epsffile {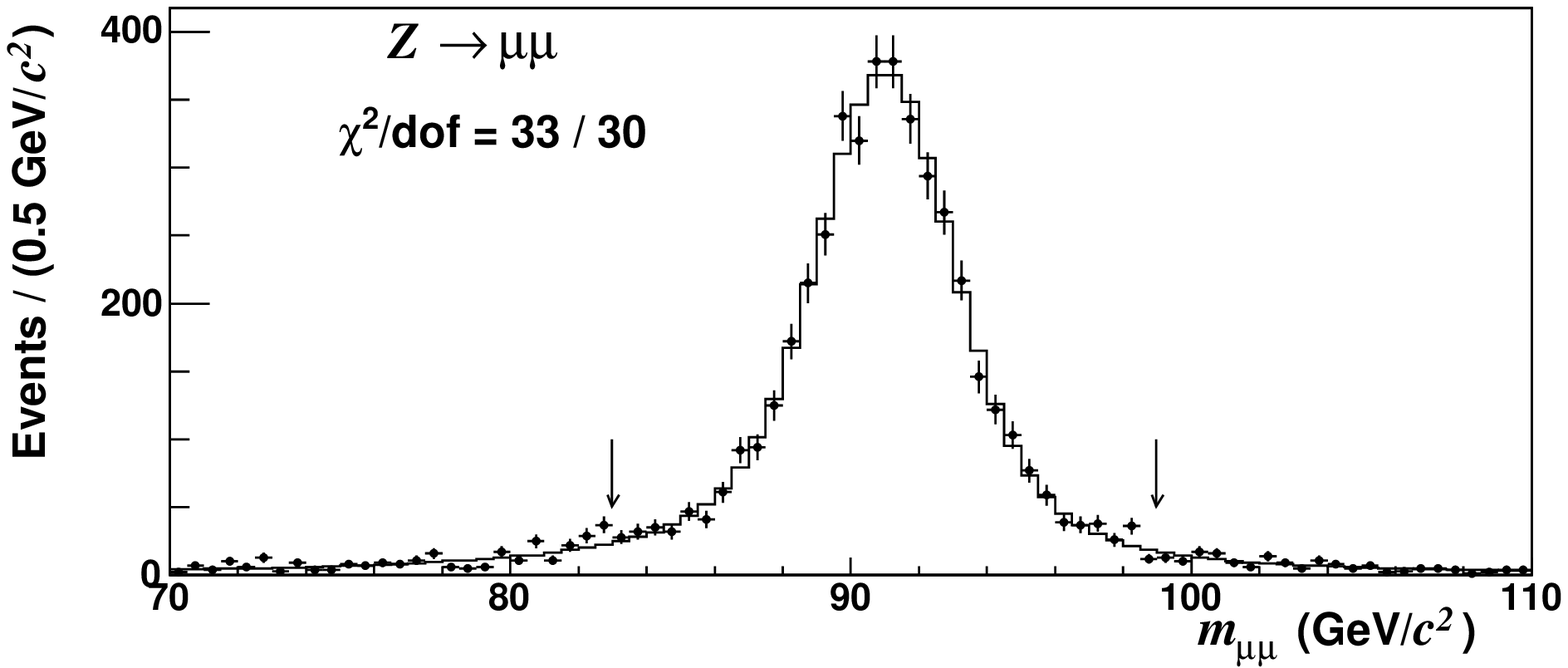}
\caption{The $Z \rightarrow \mu \mu $ (top) and $Z \rightarrow ee $ (bottom) 
 mass fits, showing the data (points) and the 
  simulation (histogram). The arrows indicate the fitting range. }
\label{z_mass_fit}
\end{center}
\end{figure}

 The tracker momentum scale is measured  
  by template-fitting the $J/\psi \rightarrow \mu \mu$ and $\Upsilon \rightarrow \mu \mu$
 mass peaks.
  The $J/\psi$ fits are performed in bins of $\left\langle 1/p_T^\ell(\mu) \right\rangle$ to measure 
 any non-linearity due to mismodelling of the ionization energy loss and other smaller effects,
 and in bins of $\left\langle \cot \theta (\mu)
 \right\rangle$ to measure the magnetic field non-uniformity. 
 To account for the observed momentum non-linearity, 
 a downward 6\% correction to the predicted ionization
 energy loss is applied in the simulation to make
 the measured $J/\psi$ mass independent of $\left\langle 1/p_T^\ell(\mu) \right\rangle$. 
 The  calibration derived from the $J/\psi$  and $ \Upsilon$ data 
 yields $M_Z = ( 91184 \pm 43_{\rm stat} )$~\mevcc\ 
 (Fig.~\ref{z_mass_fit}) from the $Z \rightarrow \mu \mu$  data, consistent with the 
 world average~\cite{langacker,lepwmass} of $( 91188 \pm 2 )$~\mevcc. 
 The systematic uncertainties  due to QED radiative corrections and
  magnetic field non-uniformity dominate the total  uncertainty of 0.02\% on 
 the combined momentum scale,  derived from the  $J/\psi , \Upsilon$
 and  $Z $ boson mass fits.

\begin{figure}[htb]
\begin{center}
\epsfysize = 4.4cm
\epsffile {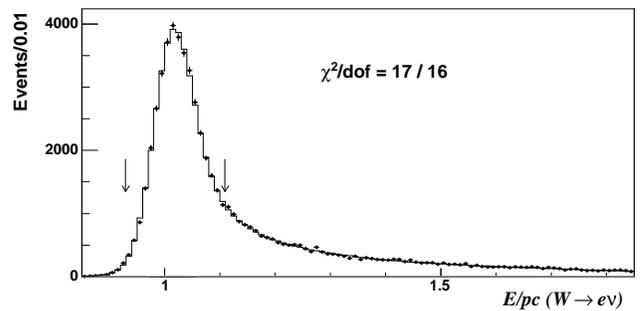}
\caption{The distribution of $E / pc$ for the $W \rightarrow e \nu$ data (points) and the 
  best-fit simulation (histogram) including the small jet background (shaded). The arrows indicate the fitting range used for the
 electron energy calibration. The jet background, which is barely visible on this scale, contributes
 a negligible uncertainty in the calibrations of the electron energy scale and the amount of radiative material.}
\label{eop}
\end{center}
\end{figure}

 We simulate the electron cluster by merging the energies of the primary electron and the proximate 
 bremsstrahlung photons and conversion electrons. The distributions
 of electron and photon  energy loss in the solenoid coil,  
 and leakage into the hadronic calorimeter, are determined using {\sc geant}~\cite{geant} 
 as a function of $E_T$. The fractional energy resolution is given by the quadrature sum of a
  sampling term (13.5\%/$\sqrt{E_T/ \rm{GeV}}$) and a constant term $\kappa = (0.89 \pm 0.15)$\% applied to the 
 cluster energy, and an additional constant term  $\kappa_\gamma = (8.3 \pm 2.2)$\% 
 applied only to the energies of bremsstrahlung photons and 
 conversion electrons. The $\kappa_\gamma$ term contributes  $\approx 1.3$\% in quadrature to the effective constant term for the inclusive electron
 sample. The distribution of the underlying event energy~\cite{prd}  in the cluster 
 is simulated. We tune $\kappa$ 
 on the width of the $E / pc$
 peak (Fig.~\ref{eop}) of the $W$ boson sample,
  and  $\kappa_\gamma$  on the width of the $Z \rightarrow ee$ mass peak when both electrons are radiative
 ($E / pc > 1.06$). 

 Given the tracker momentum calibration, we fit the   $E / pc$ peak 
 in bins of electron  $E_T$
  to determine the electron energy scale and non-linearity. The position of the $E / pc$ peak 
 is sensitive to 
 the number of radiation lengths $X_0$ ($\approx$19\%),
 due to bremsstrahlung upstream of the COT. We constrain $X_0$ by comparing the fraction of 
 electrons with
 high $E / pc$ between data and simulation.   
 Applying the $E/pc$-based energy calibration, we fit the  $Z \rightarrow ee$ mass peak and 
 measure $M_Z = ( 91190 \pm 67_{\rm stat} ) $ \mevcc\ (Fig.~\ref{z_mass_fit}),
 consistent with the world average~\cite{langacker,lepwmass}. For maximum precision, 
 the energy scales from the $W$ $E / pc$ 
 fit and the $Z \rightarrow ee$ mass fit are combined  using the 
 Best Linear Unbiased Estimate (BLUE) method~\cite{bluemethod}, with a resulting uncertainty that is 
 mostly  statistical. 

\begin{figure}[b]
\begin{center}
\epsfysize = 4.1cm
\epsffile {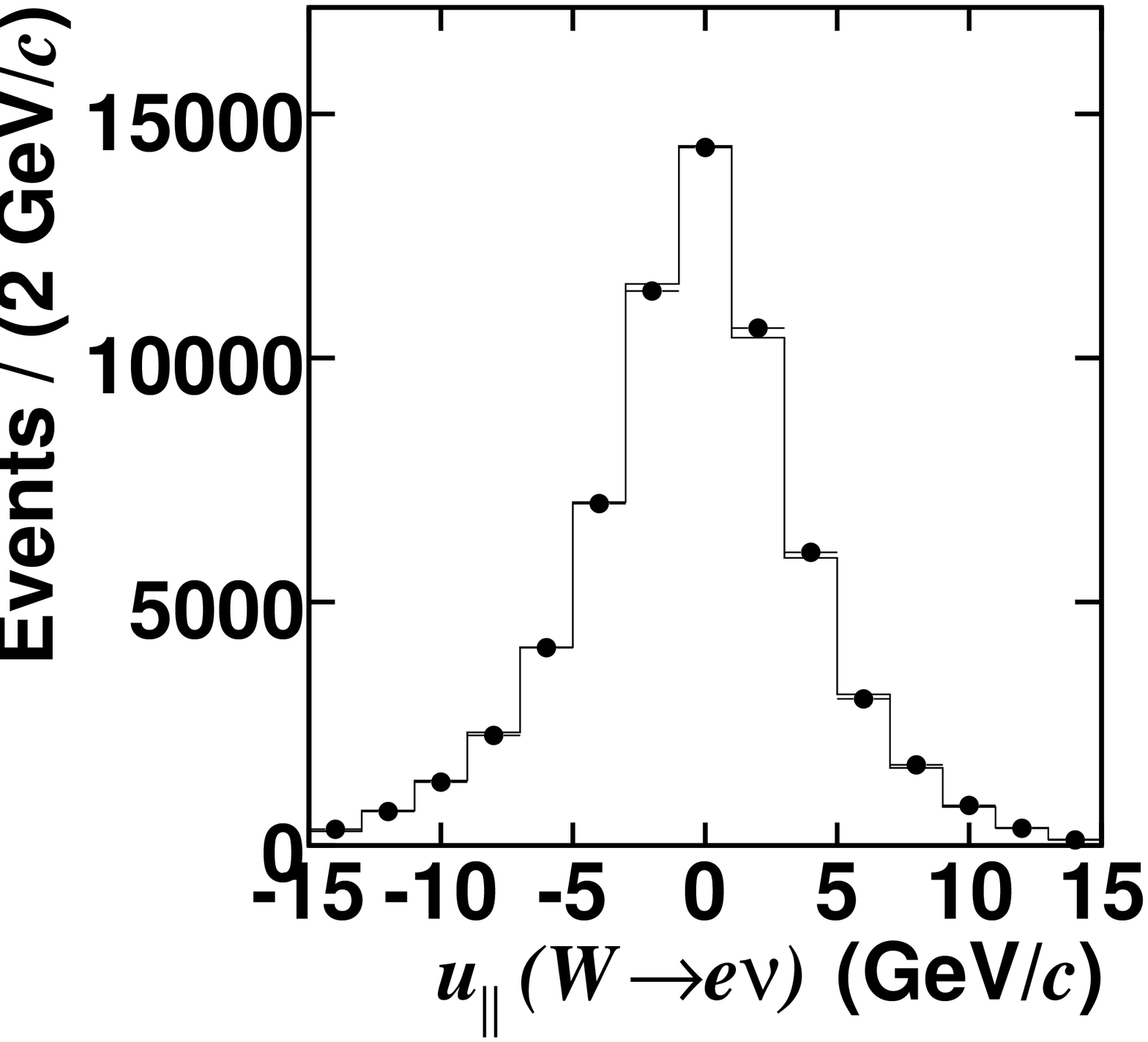}
\epsfysize = 4.1cm
\epsffile {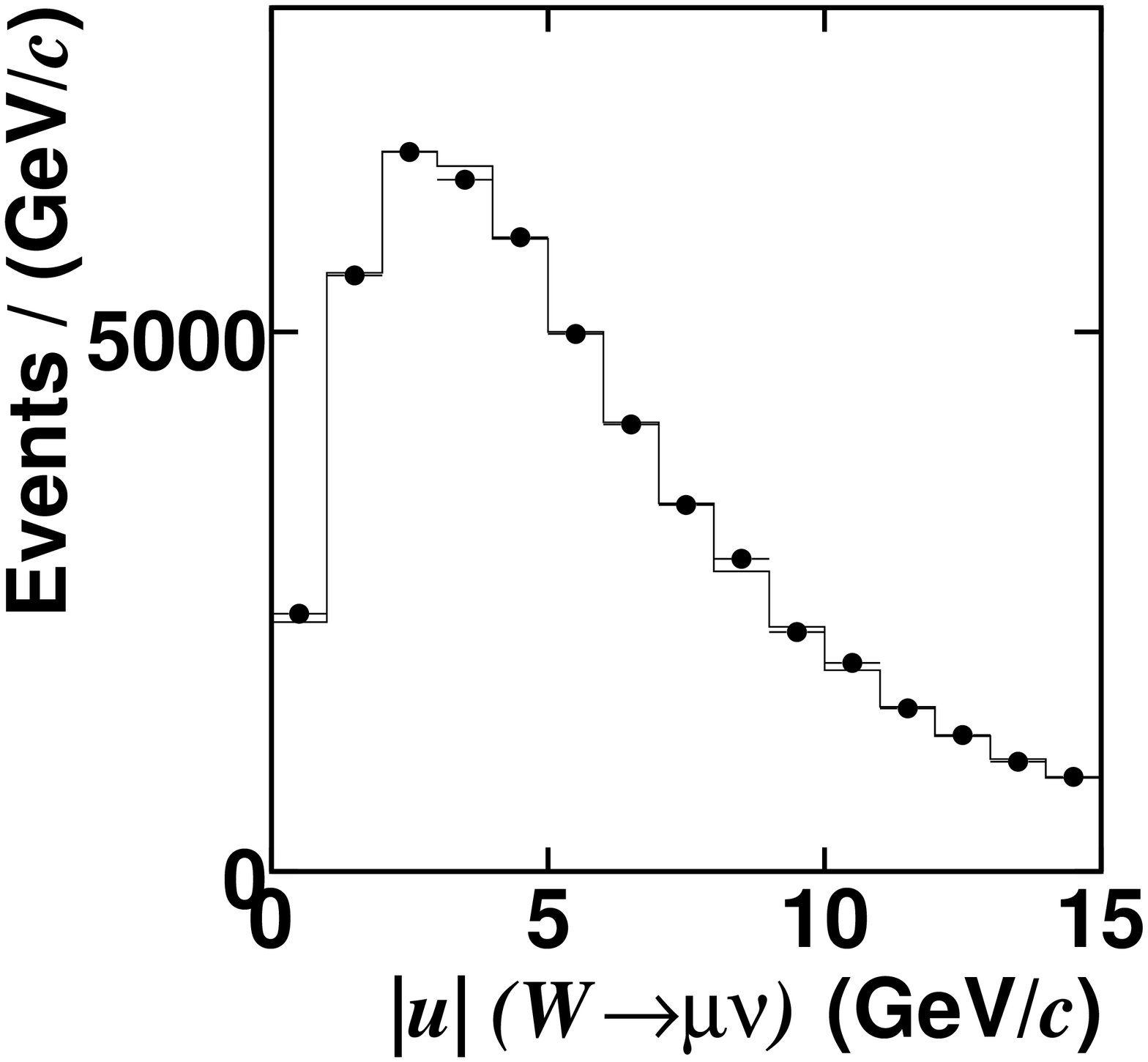}
\caption{Left:  The  $u_{||} $ distribution for the electron channel data (points) and simulation (histogram).
  Right: The $|\vec{u}| $ distribution for the muon channel.
 The mean and r.m.s. of the histograms agree
             between data and simulation, within the statistical precisions of $\approx 1$\%.
 }
\label{fig:wplots}
\end{center}
\end{figure}

 The  recoil $\vec{u}$ excludes  towers in which the lepton(s) deposit energy. The
 underlying event energy in these towers is measured from the nearby towers in $W$ boson  data,
 including its dependence on $\eta^\ell$ and $\vec{u}$. 
 The  resolution of $\vec{u}$  has  jet-like  and  underlying event components,
 with the latter modelled using data triggered on  inelastic $\bar{p} p$ interactions.  
 The recoil  parameterizations are tuned on the mean and r.m.s. of the
 $\vec{p}_T$ imbalance between the dilepton $\vec{p}_T$ and $\vec{u}$ 
 in $Z \rightarrow \ell \ell$ events.  
%A relative calibration of the hadronic response of
% the central and plug calorimeters, performed using hadron tracks,  minimizes
% the sensitivity of the recoil model to differences 
% in the phase space sampled by the selected $Z$ and $W$ boson decays.
 The lepton identification efficiency is measured as
 a function of $u_{||} =   \vec{u} \cdot \vec{p}_T^{~\ell} / p_T^{~\ell} $ 
 using the $Z \rightarrow  \ell \ell$ data, in order to model its effect 
 on the  $p_T^\ell$  and  $p_T^\nu$ distributions. 
 Cross-checks of the
 recoil model using the $W$ boson data show good agreement (Fig.~\ref{fig:wplots}). 
 
\begin{table}[htpb]
\begin{center}
\begin{tabular}{ccc}
\hline
\hline
 Distribution~  & ~ $W$ boson mass (MeV/$c^2$)~  &~ $\chi^2$/dof \\
\hline
 $m_T(e,\nu)$  &  $80493 \pm 48_{\rm stat} \pm  39_{\rm syst}$   & 86/48   \\
 $p_T^\ell(e)$        &  $80451 \pm 58_{\rm stat} \pm  45_{\rm syst}$   & 63/62   \\
 $p_T^\nu(e)$       &  $80473 \pm 57_{\rm stat} \pm  54_{\rm syst}$   & 63/62   \\
\hline
 $m_T (\mu,\nu)$ &  $80349 \pm 54_{\rm stat} \pm  27_{\rm syst}$   & 59/48   \\
 $p_T^\ell(\mu)$       &  $80321 \pm 66_{\rm stat} \pm  40_{\rm syst}$   & 72/62   \\
 $p_T^\nu (\mu)$      &  $80396 \pm 66_{\rm stat} \pm  46_{\rm syst}$   & 44/62   \\
\hline
\hline
\end{tabular}
\caption{Fit results and uncertainties
 for $M_W$. 
 The fit windows are $65-90$~\gevcc for the $m_T$ fit
and $32-48$~\gevc for the $p_T^\ell$ and $p_T^\nu$ fits.
 The $\chi^2$ of the fit is computed using the expected 
 statistical errors on the data points. }
\label{tab:fitResults}
\end{center}
\end{table}

 Backgrounds in the $W$ boson candidate samples arise  
 from  misidentified jets containing high-$p_T$ tracks and EM clusters,
  $Z \rightarrow  \ell \ell$ where a lepton is not reconstructed and mimics
 a neutrino, 
 $W \rightarrow \tau \nu$, $\pi/K$ decays in flight (DIF),  
 and cosmic rays.
 Jet, DIF, and cosmic ray backgrounds are estimated
 from the  data 
%by fitting $W$ boson distributions to a sum of expected signal (obtained from simulation or 
% $Z \rightarrow \ell \ell$ data) and background-enriched distributions (obtained using leptons and/or tracks of poor
% quality), and are found 
to be  less than 0.5\% combined. 
 The  $W \rightarrow \tau \nu$ background is $0.9$\% ($0.9$\%), and the $Z \rightarrow  \ell \ell$
 background  is 6.6\% (0.24\%) in the muon (electron) channel, as
 estimated using a detailed {\sc geant}-based detector simulation. 
The background shapes are obtained using simulation and data-derived  distributions.

\begin{figure}[htb]
\begin{center}
\epsfysize = 4.5cm
\epsffile {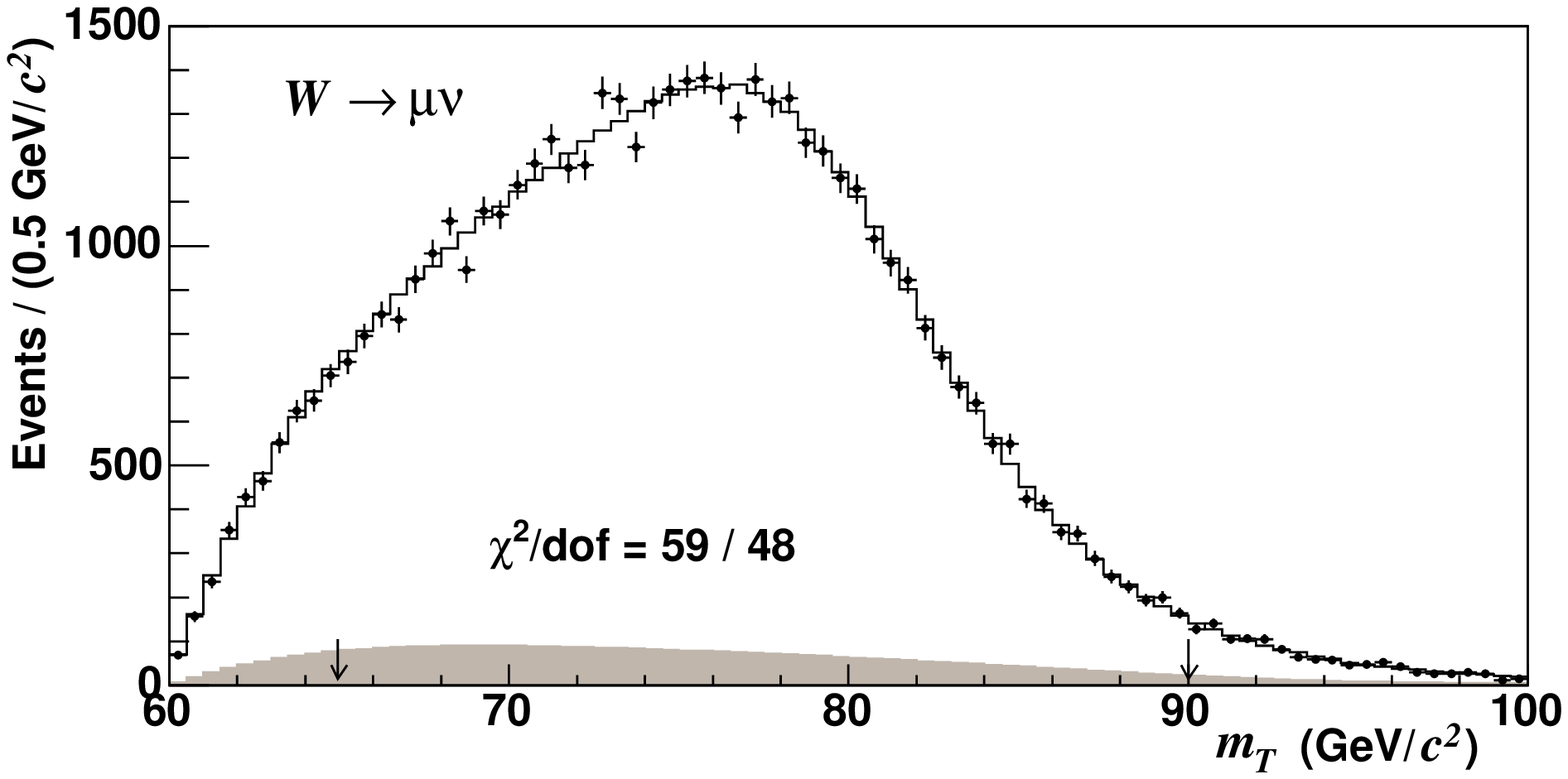}
\epsfysize = 4.5cm
\epsffile {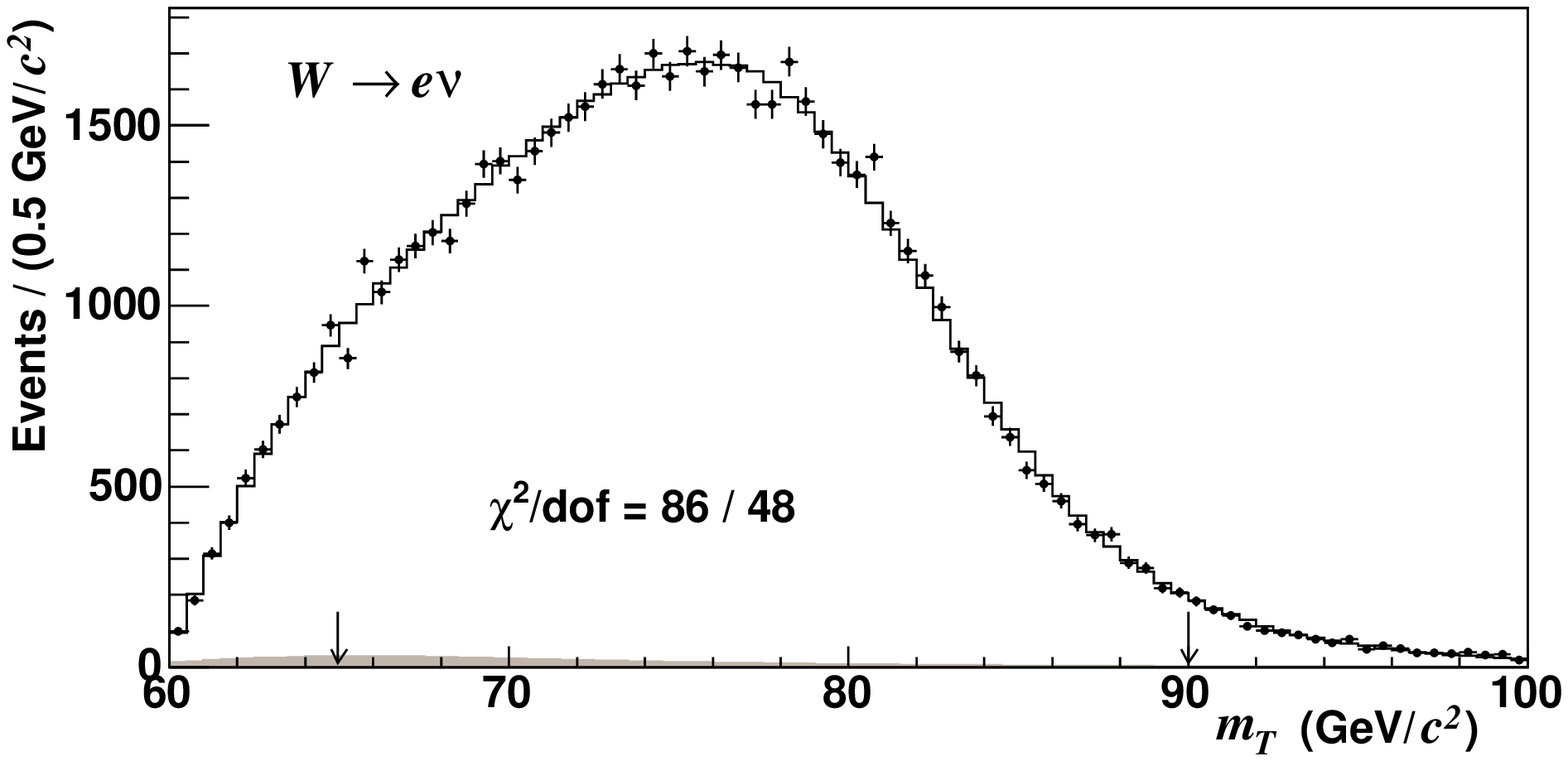}
\caption{ The $m_T $ distribution of the data (points)
 and the 
best-fit simulation template (histogram) including backgrounds (shaded),  for muons (top) and electrons (bottom). 
 The arrows indicate the fitting
 range. The $\chi^2$/dof for the electron channel distribution receives large contributions from
 a few bins near 65~GeV/$c^2$, which do not bias the mass fit.
}
\label{w_mass_fit}
\end{center}
\end{figure}

\begin{table}[*h]
\begin{center}
\begin{tabular}{cccc}
\hline
\hline
Systematic    & $W \rightarrow e \nu$ & $W \rightarrow \mu \nu$ & Common  \\ [0.05in]
\hline
$p_T(W)$ model                   & \sysgtwomt\   & \sysgtwomt\  & \sysgtwomt\     \\
QED radiation             &    \sysqedmte\     & \sysqedmtm\  & \sysqedmte\    \\
Parton distributions      &    11     & 11    &    11      \\
Lepton energy scale              &    \sysescale\ & \syspscale\    &  \syspscale\        \\
Lepton energy resolution         & \syslepresmte\  & \syslepresmtm\  &     0      \\
Recoil energy scale              &  \sysrecrespmt\  &   \sysrecrespmt\  &  \sysrecrespmt\ \\
Recoil energy resolution         &  \sysrecresolmt\ &  \sysrecresolmt\  &   \sysrecresolmt\   \\
$u_{||}$ efficiency       &   3  &  1  &     0   \\
Lepton removal            &    8     &  5    &     5  \\
Backgrounds               &    8     & \sysmbkgmt\    &     0       \\
\hline
Total systematic                    &    39     & 27    &    26  \\
\hline
Total uncertainty                   &    62     & 60    &    26  \\
\hline
\hline
\end{tabular}
\caption{Systematic and total uncertainties in  \mevcc\ for the $m_T$ fits, which are the most precise.
  The last column shows 
 the correlated uncertainties. }
\label{tab:systematics}
\end{center}
\end{table}

Table~\ref{tab:fitResults} shows the fit results 
 from the $m_T$ (Fig.~\ref{w_mass_fit}), 
 $p_T^\ell$, and $p_T^\nu$ distributions. These fits are partially uncorrelated and have
 different systematic uncertainties, thus providing an important cross-check. 
 The fit values were hidden during analysis by adding an unknown offset in the range [-100, 100] \mevcc. 
 The systematic uncertainties (Table~\ref{tab:systematics}) were evaluated by fitting MC events to 
 propagate 
 the  analysis 
 parameter uncertainties to $M_W$.

The consistency of the fit results (Table~\ref{tab:fitResults}) obtained from the different  distributions
shows that the $W$ boson production, decay, and the hadronic recoil are well-modeled.
The statistical 
 correlations (evaluated using ensembles of MC events) 
 between the $m_T$ and $p_T^\ell$ ($p_T^\nu$) fit values is 69\% (68\%), and  between
 the $p_T^\ell$ and $p_T^\nu$ fit values is 27\%.   
 We numerically combine (using the BLUE method) the 
  six individually fitted $M_W$ values, including their correlations, to obtain 
 $M_W = (80413 \pm 34_{\rm stat} \pm 34_{\rm syst}) $ \mevcc, with 
 $\chi^2 / {\rm dof} = 4.8 / 5$. 
 The $m_T$, $p_T^\ell$ and $p_T^\nu$ fits in the electron (muon) channel 
 contribute  weights of 
 32.3\% (47.7\%), 8.9\% (3.4\%) and 6.8\% (0.9\%) respectively. 
 This establishes an {\it a-priori}
 procedure to incorporate 
 the information from individual fits. The muon (electron) channel  alone  yields 
 $M_W = (80352 \pm 60) $ \mevcc\ ($M_W = (80477 \pm 62) $ \mevcc) 
  with   $\chi^2 / {\rm dof} = 1.4 / 2 ~(0.8 / 2)$. The  $m_T$ ($p_T^\ell$, $p_T^\nu$) fit
 results from the muon and electron channels are consistent with a probability of 7\% (18\%, 43\%), 
 taking into account their correlations. 

In conclusion, we report the first measurement of the $W$ boson mass from Run II 
 of the Tevatron. We measure 
 $M_W  = (80413 \pm 48)$ \mevcc, the most precise single measurement to date, and we
 update the world average~\cite{lepwmass} to   $M_W  = (80398 \pm 25)$ \mevcc.
This analysis significantly improves in precision over
      previous Tevatron measurements, not only through the increased integrated
      luminosity but also through improved analysis techniques and 
 understanding of systematic uncertainties. As many  simulation
  parameters are constrained by data control samples, their uncertainties
 are statistical in nature and are expected to be reduced with more data. 
Inclusion of our result in the global electroweak fit~\cite{lepwmass,prd} reduces the
 predicted mass of the SM Higgs boson by 6~\gevcc and decreases its range to
$m_H = 76^{+33}_{-24}$~\gevcc. 

We thank the Fermilab staff and the technical staffs of the participating institutions 
for their vital contributions. This work was supported by the 
U.S. Department of Energy and National Science Foundation; the Italian 
Istituto Nazionale di Fisica Nucleare; the Ministry of Education, Culture, 
Sports, Science and Technology of Japan; the Natural Sciences and Engineering 
Research Council of Canada; the National Science Council of the Republic of 
China; the Swiss National Science Foundation; the A.P. Sloan Foundation; the 
Bundesministerium f\"ur Bildung und Forschung, Germany; the Korean Science and 
Engineering Foundation and the Korean Research Foundation; the Science and Technology 
Facilities Council and the Royal Society, UK; the Institut National de Physique 
Nucleaire et Physique des Particules/CNRS; the Russian 
Foundation for Basic Research; the Comisi\'on Interministerial de Ciencia y 
Tecnolog\'{\i}a, Spain; the European Community's Human Potential Programme; 
the European Commission under the Marie Curie Programme; the Slovak R\&D Agency; 
and the Academy of Finland. 


\begin{references}
\bibitem{coords}
 Pseudorapidity is defined as 
 $\eta = -\ln[\tan(\theta/2)$], where $\theta$ is the polar angle from the beam axis.  Energy (momentum)  transverse to
  the beam is denoted
as $E_T$ ($p_T$).
\end{references}
\end{document}